\documentclass[12pt]{article}
\usepackage{amsmath}
\usepackage{amssymb,palatino,theorem}
\usepackage{graphicx}
\usepackage{psfig}

\newcommand{\ls}[1]
    {\dimen0=\fontdimen6\the\font \lineskip=#1\dimen0
\advance\lineskip.5\fontdimen5\the\font \advance\lineskip-\dimen0
\lineskiplimit=.9\lineskip \baselineskip=\lineskip
\advance\baselineskip\dimen0 \normallineskip\lineskip
\normallineskiplimit\lineskiplimit
\normalbaselineskip\baselineskip \ignorespaces }

\newtheorem{proposition}{Proposition}

\makeatletter
\def\@begintheorem#1#2{\it \trivlist \item[\hskip \labelsep{\bf #1\
#2.}]} \makeatother

\newcommand{\be}{\begin{equation}}
\newcommand{\ee}{\end{equation}}
\newcommand{\bea}{\begin{eqnarray}}
\newcommand{\eea}{\end{eqnarray}}

\newcommand{\beq}[1]{\begin{equation}\label{#1}}
\newcommand{\eeq}{\end{equation}}

\newcommand{\beqn}[1]{\begin{eqnarray}\label{#1}}
\newcommand{\eeqn}{\end{eqnarray}}

\newcommand{\beaa}{\begin{eqnarray*}}
\newcommand{\eeaa}{\end{eqnarray*}}
\def\E{{\mathbb E}}

\def\SNR    {\mbox{\scriptsize\sf SNR}}

\def\MMSE    {\mbox{\scriptsize\sf MMSE}}

\def \det       {{\rm det}}

\def \arg       {{\rm arg}}

\def \nT        {n_\mathrm{\scriptscriptstyle T}}
\def \nR        {n_\mathrm{\scriptscriptstyle R}}

\def\non{\nonumber\\}

\def\Hm{{\bf H}}

\def\AbvGT #1#2{\lower2pt\vbox{\baselineskip0pt \lineskip-.5pt%
         \halign{$#1 ##$\cr #2\crcr >\cr}}}

\def\fd {f_{\rm \scriptscriptstyle D}}

\def\complex{\mathop{\raise .45ex\hbox{${\bf\scriptstyle{|}}$}
      \kern -0.40em {\rm \textstyle{C}}}\nolimits}
\def\hilbert{\mathop{\raise .21ex\hbox{$\bigcirc$}}\kern -1.005em
{\rm\textstyle{H}}} 

\def\PARstart#1#2{\begingroup\def\par{\endgraf\endgroup\lineskiplimit=0pt}
    \setbox2=\hbox{\uppercase{#2} }\newdimen\tmpht \tmpht \ht2
    \advance\tmpht by \baselineskip\font\hhuge=cmr10 at \tmpht
    \setbox1=\hbox{{\hhuge #1}}
    \count7=\tmpht \count8=\ht1\divide\count8 by 1000 \divide\count7 by\count8
    \tmpht=.001\tmpht\multiply\tmpht by \count7\font\hhuge=cmr10 at \tmpht
    \setbox1=\hbox{{\hhuge #1}} \noindent \hangindent1.05\wd1
    \hangafter=-2 {\hskip-\hangindent \lower1\ht1\hbox{\raise1.0\ht2\copy1}%
    \kern-0\wd1}\copy2\lineskiplimit=-1000pt}

\def\squarebox#1{\hbox to #1{\hfill\vbox to #1{\vfill}}}

\title{\bf Optimum Pilot Overhead in Wireless Communication: A Unified
Treatment of Continuous and Block-Fading Channels}

\author{Nihar Jindal\thanks{Nihar Jindal is with the University of Minnesota,
Minneapolis, MN 55455, USA. His work was partially conducted during a visit to UPF under the sponsorship of Project TEC2006-01428.} and Angel Lozano\thanks{Angel Lozano is with Universitat Pompeu Fabra (UPF), 08018 Barcelona, Spain.
His work is supported by the projects CONSOLIDER-INGENIO 2010 CSD2008-00010 "COMONSENS" and IRG-224755 "NetMIMO".}}

\begin{document}

\maketitle

\begin{abstract}
The optimization of the pilot overhead in single-user wireless fading channels is investigated, and the dependence of this
overhead on various system parameters of interest (e.g., fading rate, signal-to-noise ratio) is quantified.
The achievable pilot-based spectral efficiency is expanded with respect to the fading rate about the no-fading point, which leads
to an accurate order expansion for the pilot overhead.  This expansion identifies that the pilot overhead, as well
as the spectral efficiency penalty with respect to a reference system with genie-aided CSI (channel state information) at the receiver,
depend on the square root of the normalized Doppler frequency.  Furthermore, it is shown that the widely-used block fading
model is only a special case of more accurate continuous fading models in terms of the achievable
pilot-based spectral efficiency, and that the overhead optimization for multiantenna systems is effectively the
same as for single-antenna systems with the normalized Doppler frequency multiplied by the number of transmit antennas.
\end{abstract}


\eject

\section{Introduction}

\begin{figure}
  \centering
 \includegraphics[width=0.6\linewidth]{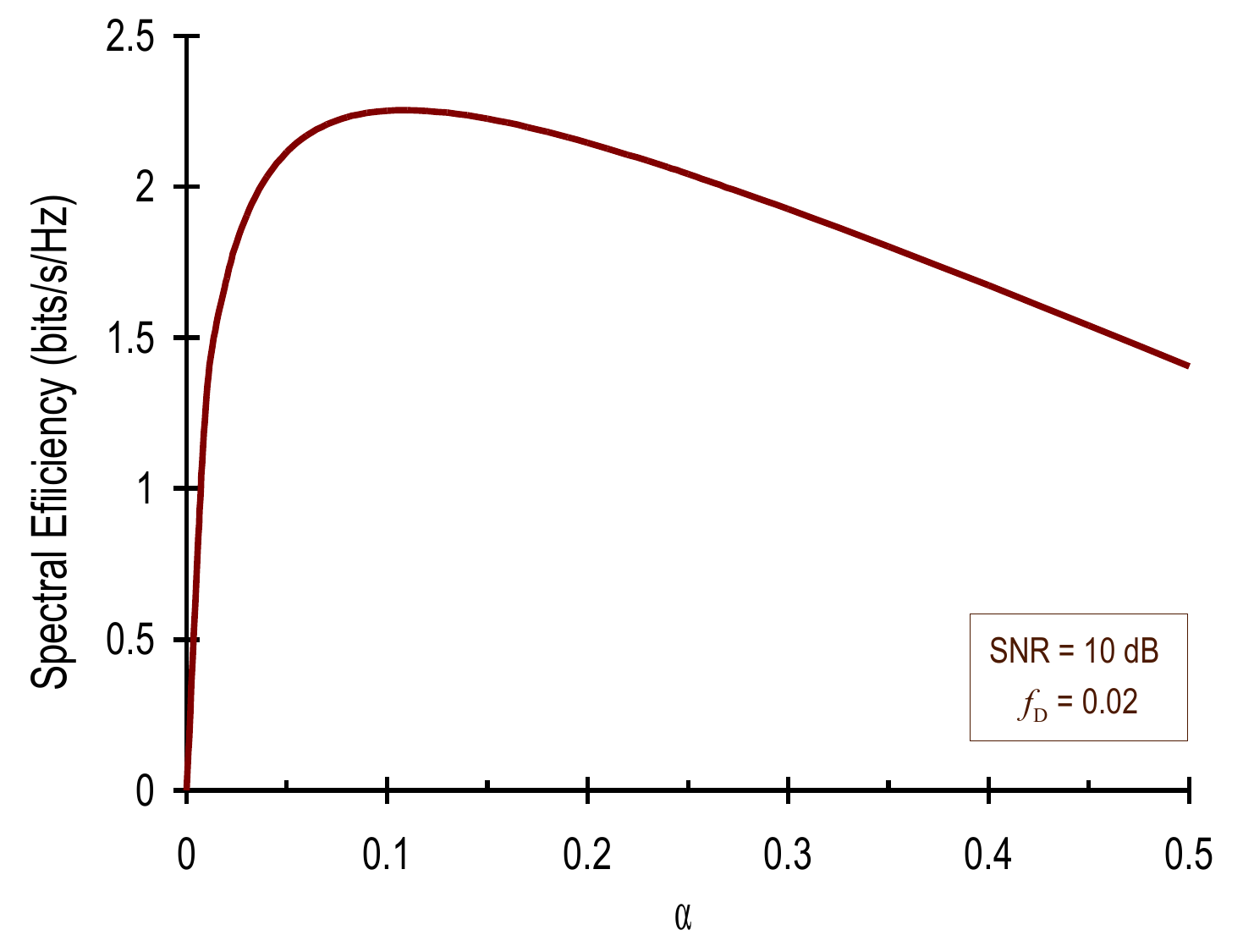}
  \caption{Spectral efficiency as function of the pilot overhead, $\alpha$, for $\SNR = 10$ dB. The Doppler spectrum is Clarke-Jakes with a maximum
  normalized frequency $\fd=0.02$ corresponding, for instance, to $100$ Km/h in a WiMAX system.}
  \label{intro}
\end{figure}

Most wireless communication systems perform coherent data detection with the assistance of pilot signals (a.k.a. reference signals
or training sequences) that are inserted periodically \cite{Cavers91,Tong04}. The receiver typically performs channel estimation
on the basis of the received pilot symbols, and then applies standard coherent detection while treating the channel
estimate as if it were the true channel. When such an approach is taken and Gaussian inputs are used, the channel estimation error
effectively introduces additional Gaussian noise \cite{medard,amos-shamai}.
This leads to a non-trivial tradeoff: increasing the fraction of symbols that serve as pilots improves the
channel estimation quality and thus decreases this additional noise, but also decreases the fraction of symbols that can carry data.
To illustrate the importance of this tradeoff, Fig. \ref{intro} depicts the spectral efficiency as function of the pilot overhead
(cf. Section \ref{Kobe} for details) for some standard channel conditions.
Clearly, an incorrect overhead can greatly diminish the achievable spectral efficiency.

Although this optimization is critical and has been extensively studied in the literature
\cite{medard}--\nocite{amos-shamai,Zheng02,hassibi,MaYangGiannakis,zheng-tse-medard,Furrer07,Baltersee01,Ohno02,DengHaimovich07}\cite{lozano08},
on the basis of both the simplified block-fading model as well as the more accurate continuous-fading model, such
optimization must be solved numerically except for one known special
case.\footnote{A closed-form solution for the optimal overhead when the power of the pilot symbols can be boosted in a block-fading
channel model is derived in \cite{hassibi}.}
Indeed, other than some low- and high-power asymptotes,
no explicit expressions are available to identify the optimum overhead or
to assess how it depends on the various parameters of interest (velocity, power, etc).

In this paper, we circumvent this difficulty by studying the overhead optimization in the limiting regime of slow fading.
More precisely, by expanding the spectral efficiency around the perfect-CSI point, i.e., for
small fading rates, the optimization can be tackled and a useful expansion (in terms of the fading rate) for the optimum pilot
overhead is obtained.  The key insights reached in the paper are as follows:

\begin{itemize}

\item In terms of the spectral efficiency
achievable with channel estimate-based decoding,
block-fading is simply a special case of continuous (symbol-by-symbol) fading.

\item The optimal pilot overhead scales with the square
root of the Doppler frequency; this result holds regardless of whether pilot power boosting is allowed.\footnote{To the best of our
knowledge, this square-root dependence was first identified in the
context of a different (and weaker) lower bound for the multiantenna broadcast channel in \cite{kobayashi08}.}

\item The spectral efficiency penalty w.r.t. the perfect-CSI capacity also
scales with the square-root of the fading rate.

\item The pilot overhead optimization for multiantenna transmission
is essentially the same as the optimization for single-antenna
transmission except with the true Doppler frequency multiplied by the number of transmit antennas.

\end{itemize}

\section{Preliminaries}
\label{preliminaries}

\subsection{Channel Model}
\label{sec:channel_model}

Consider a discrete-time frequency-flat scalar fading channel $H(k)$ where $k$ is the time index.
(The extension to multiantenna channels is considered in Section \ref{mimo}.)
Pilot symbols are inserted periodically in the transmission \cite{Dong04} and the fraction thereof is denoted by $\alpha$, i.e., one in
every $1/\alpha$ symbols is a pilot while the rest are data. Moreover, $\alpha \geq \alpha_{\rm min}$ where $\alpha_{\rm min}$ is
established later in this section.

Let $\mathcal{D}$ denote the set of time indices corresponding to data symbols.
For $k \in \mathcal{D}$,
\be
\label{channelmodel}
Y(k)= H(k) \sqrt{P} X(k) + N(k)
\ee
where the transmitted signal, $X(k)$, is a sequence of IID (independent identically distributed)
complex Gaussian random variables with zero mean and unit variance that we indicate by $X \sim\mathcal{N}_{\mathbb{C}}(0,1)$.
The additive noise is $N \sim\mathcal{N}_{\mathbb{C}}(0,N_0)$ and we define $\SNR=P / N_0$.

For $k \notin \mathcal{D}$, unit-amplitude pilots are transmitted and thus
\be
Y(k)= H(k) \sqrt{P} + N(k) .
\ee
Notice that pilot symbols and data symbols have the same average power.
In Section~\ref{milito}, we shall lift this constraint allowing for power-boosted pilots.

\subsubsection{Block Fading}

In the popular block-fading model, the channel is drawn as $H \sim\mathcal{N}_{\mathbb{C}}(0,1)$ at the beginning of each
block and it then remains constant for the $n_{\sf b}$ symbols composing the block. This process is repeated for every block in an IID fashion.

In order for the receiver to estimate the channel, at least one pilot symbol must be inserted within each block.  If
$n_{\sf p}$ represents the number of pilot symbols in every block, then
\be
\alpha = \frac{n_{\sf p}}{n_{\sf b}}
\ee
and clearly $\alpha_{\rm min} = 1/n_{\sf b}$.

\subsubsection{Continuous Fading}

In this model, $H(k)$ is a discrete-time complex Gaussian stationary\footnote{The
block-fading model, in contrast, is not stationary but only
cyclostationary.} random process,
with an absolutely continuous
spectral distribution function whose derivative is the Doppler
spectrum $S_{H}(\nu)$, $-1/2\leq \nu \leq 1/2$. It follows that the
channel is ergodic.

The discrete-time process $H(k)$ is derived from an underlying continuous-time fading process whose Doppler spectrum is $S(f)$.
We consider bandlimited processes such that
\be
\label{lovells}
\left\{ \begin{array}{ll}
          S(f)>0, & \quad |f|\leq f_{\sf m} \\
          S(f)=0, & \quad |f|> f_{\sf m}
        \end{array}
        \right.
\ee
User motion generally results in $f_{\sf m}=v/\lambda$
where $v$ is the velocity and $\lambda$ is the carrier wavelength.
(Higher values for $f_{\sf m}$ may result if the reflectors are also in motion or if multiple reflexions take place.)

Denoting by $T$ the symbol period and by $\Pi(\cdot)$ the Fourier
transform of the transmission pulse shape, the spectrum of the
discrete-time and continuous-time processes are related according to
\be \label{grammy}
S_{H}(\nu)=\frac{1}{T} \, S\left(\frac{\nu}{T}\right) \Pi^2(\nu). \ee
As a result, the discrete-time spectrum is nonzero only for $|\nu|\leq f_{\sf m} T$.\footnote{Note that (\ref{grammy}) implies a matched-filter
front-end at the receiver. This entails no loss of optimality if $f_{\sf m} \ll 1/T$, a premise usually
satisfied, and the smooth pulse shaping $\Pi(\cdot)$ can thereby be disregarded altogether.}
For notational convenience, we therefore define a normalized Doppler $\fd = f_{\sf m} T$.

To ensure that the decimated
channel observed through the pilot transmissions has an unaliased spectrum, it is necessary that
\be \label{cafe}
\alpha_{\sf min} = 2 \fd .\ee On account of its bandlimited nature, the channel is a nonregular fading process \cite{Doob}.
For simplicity we further consider $S_H(\cdot)$ to be strictly positive within $\pm \fd$.\footnote{This premise
can be easily removed by simply restricting all the integrals in the paper to the set of frequencies
where $S_H(\nu)>0$, rather than to the entire interval $\pm \fd$.}  In order to remain consistent with earlier definitions of signal
and noise power, only unit-power processes are considered.

Two important spectra are the Clarke-Jakes \cite{jakes}
\be
\label{cj}
S_H(\nu)= \frac{1}{\pi \sqrt{\fd^2-\nu^2}}
\ee
and the rectangular
\be \label{uniform} S_H(\nu) = \left\{ \begin{array}{ll}
          1/(2 \fd) & \quad |\nu|\leq \fd \\
          0 & \quad |\nu|> \fd .
        \end{array}
        \right.
\ee

We will later find it useful to express the Doppler spectrum as
\be
\label{eze}
S_H(\nu) = \frac{1}{\fd} \, \tilde{S}_H \! \left( \frac{\nu}{\fd} \right)
\ee
where $\tilde{S}_H(\cdot)$ is a normalized spectral shape bandlimited to $\pm 1$.
For the Clarke-Jakes spectrum in (\ref{cj}), for instance, the spectral shape is
\be
\tilde{S}_H (\nu) = \frac{1}{\pi \sqrt{1-\nu^2}}
\ee
while, for the rectangular spectrum in (\ref{uniform}), the spectral shape is
\be
\tilde{S}_H(\nu) = \left\{ \begin{array}{ll}
          1/2 & \quad |\nu|\leq 1 \\
          0 & \quad |\nu|> 1 .
        \end{array}
        \right.
\ee

%

\subsection{Perfect CSI}

With perfect CSI at the receiver, i.e., assuming a genie provides
the receiver with $H(k)$, there is no
need for pilot symbols ($\alpha=0$). The capacity in bits/s/Hz is then \cite{wcylee,ozarow}
\bea
\label{baste}
C(\SNR) & = &
 E \left[ \log_2 \left(1+ \SNR \, |H|^2 \right) \right] \\
& = &  \log_2(e) \; e^{1/\SNR} E_1 \! \left( \frac{1}{\SNR} \right)
\label{baste2}
\eea
with $E_q(\cdot)$ the exponential integral of order $q$,
\be
\label{exponentialintegral}
E_q(\zeta)=\int_1^{\infty} t^{-q} e^{-\zeta t} dt .
\ee

The first derivative of $C(\cdot)$ can be conveniently expressed
as a function of $C(\cdot)$ via
\be
\dot{C}(\SNR) = \frac{1}{\SNR} \left( \log_2 e - \frac{C(\SNR)}{\SNR} \right) .
\ee
In turn, the second derivative can be expressed as function of $C(\cdot)$ and $\dot{C}(\cdot)$ as
\be
\ddot{C}(\SNR)=-\frac{1}{\SNR^2} \left[ \log_2 e + \dot{C}(\SNR) - 2 \frac{C(\SNR)}{\SNR} \right] .
\ee

\section{Pilot-Assisted Detection}
\label{Kobe}

In pilot-assisted communication, 
decoding must be conducted on the basis of the channel outputs (data and pilots) alone, without the
assistance of genie-provided channel realizations. In this case, the
maximum spectral efficiency that can be achieved reliably is the mutual information
between the data inputs and the
outputs (data and pilots). This mutual information equals
 \be \label{reading} \lim_{K
\rightarrow \infty} \frac{1}{K} \;
  I \! \left( \underbrace{ \{X(k)\}_{k=0}^{K-1} ; \{Y(k)\}_{k=0}^{K-1} }_{k \in \mathcal{D}}
 | \underbrace{ \{Y(k)\}_{k=0}^{K-1} }_{k \notin \mathcal{D}}
 \right)
\ee
where $K$ signifies the blocklength in symbols.
Achieving (\ref{reading}), for which there is no known simplified
expression, generally requires joint data decoding and channel
estimation.

Contemporary wireless systems take the lower complexity, albeit
suboptimal, approach of first estimating the channel for each
data symbol---based exclusively upon all received pilot symbols---and then
performing nearest-neighbor decoding using these channel estimates as if they were correct.
This is an instance of mismatched decoding \cite{czisar-book}.
If we express the channel as $H(k) = \hat{H}(k) + \tilde{H}(k)$ where $\hat{H}(k)$
denotes the minimum mean-square error estimate of $H(k)$, the received symbol can be
re-written as \be \label{eff-channel} Y(k)=
\hat{H}(k) \sqrt{P} \, X(k) +
\tilde{H}(k) \sqrt{P} \, X(k) + N(k). \ee
Performing nearest-neighbor decoding as described above\footnote{More specifically, the decoder finds the
codeword $[X(1),\ldots, X(K)]$ that minimizes the distance metric
$\sum_{k=1}^K | Y(k) - \sqrt{P} \hat{H}(k)  X(k) |^2$.}
has been shown
to have the effect of making the term $\tilde{H}(k)
\sqrt{P} \, X(k)$ appear as an additional source of
independent Gaussian noise \cite{amos-shamai}. With that,
the spectral efficiency becomes \cite{medard}--\nocite{Zheng02,hassibi,Baltersee01}\cite{Ohno02}
 \be \label{mallorca} \bar{\mathcal{I}}(\SNR,
\alpha) = (1-\alpha) \, C(\SNR_{\sf eff}) \ee with \be
\label{tertulia} \SNR_{\sf eff} = \frac{\SNR \, (1-\MMSE)}{1+\SNR
\cdot \MMSE} \ee where $\MMSE = E \left[ |\tilde{H}|^2 \right]$.
Although not shown explicitly, $\MMSE$ and $\SNR_{\sf eff}$ are
functions of $\SNR$, $\alpha$ and the underlying fading model.

In addition to representing the maximum spectral efficiency achievable
with Gaussian codebooks and channel-estimate-based nearest-neighbor decoding, $\bar{\mathcal{I}}(\cdot)$ is also
a lower bound to (\ref{reading}). Because of this double significance, the maximization of $\bar{\mathcal{I}}(\cdot)$
over $\alpha$
\be \label{mallorca2} \bar{\mathcal{I}}^\star(\SNR) =
\max_{\alpha_{\rm min} \leq \alpha \leq 1} \bar{\mathcal{I}}(\SNR,
\alpha)
\ee
and especially the argument of such maximization, $\alpha^\star$, are the focal points of this paper.


The expressions in (\ref{mallorca}) and
(\ref{mallorca2}) apply to both block and continuous fading, and
 these settings differ only in how $\MMSE$ behaves as a function
of $\alpha$ and $\SNR$.

In block fading, $n_{\sf p}$ pilot symbols are used to
estimate the channel in each block and thus \cite{hassibi} \be
\label{mmse-block} \MMSE =
\frac{1}{1 + \alpha \, n_{\sf b} \SNR}. \ee

%
%
%


For continuous fading, on the other hand \cite{Tong04,Ohno02}
\bea
\label{newmexico} \MMSE &=&  1- \int_{-\fd}^{+\fd}
\frac{\SNR \, S_H^2(\nu)}{1/ \alpha +\SNR \, S_H(\nu)} \, d\nu \\
&=& 1-\int_1^{+1} \frac{\tilde{S}_H^2(\xi)}{\frac{\fd}{\alpha \, \SNR} +\tilde{S}_H(\xi)} \, d\xi
\label{northwest}
\eea
where the latter is derived based upon the spectral shape definition in (\ref{eze}).

For the Clarke-Jakes spectrum, (\ref{newmexico}) can be computed in closed form as
\cite{lozano08} \be \label{kobe} \MMSE = 1 - \frac{{\rm arctanh}
\sqrt{1-\left(\frac{ \alpha \, \scriptscriptstyle \sf
SNR}{\pi \fd}\right)^2}} {\frac{\pi}{2} \sqrt{\left(
 \frac{\pi \fd}{\alpha \, {\scriptscriptstyle \sf
SNR}} \right)^2-1}} \ee
while, for the rectangular spectrum \cite{Ohno02}
\be \label{mmse-lpf} \MMSE
= \frac{1}{1 + \frac{\alpha}{2 \fd} \SNR}. \ee

Comparing (\ref{mmse-block}) with (\ref{mmse-lpf}),
the block-fading model is seen to yield the same $\MMSE$ as a continuous fading model with a rectangular spectrum where
\be
\label{veggieworld}
\fd = \frac{1}{2 n_{\sf b} }. \ee
Because $\bar{\mathcal{I}}(\cdot)$ depends on the fading model only through
$\MMSE$, this further implies equivalence in terms of spectral efficiency.  Thus, for the remainder of the paper we shall
consider only continuous fading while keeping in mind that block-fading corresponds to the special case of a rectangular spectrum with (\ref{veggieworld}).

%
%

\section{Pilot Overhead Optimization}
\label{petra}

The optimization in (\ref{mallorca2}) does not yield an analytical solution, even for the simplest of fading
models, and therefore it must be computed numerically.\footnote{Such numerical computation is further complicated
by the fact that for most spectra other than Clarke-Jakes and
rectangular, a closed-form solution for $\MMSE$ does not even exist.}
In this section, we circumvent this difficulty by appropriately
expanding the objective function $\bar{\mathcal{I}}(\cdot)$. This leads to a simple expression that cleanly illustrates
the dependence of $\alpha^*$ and $\bar{\mathcal{I}}^*$ on the parameters of interest.

In particular, we shall expand (\ref{mallorca}) with respect to $\fd$ while keeping the shape of the Doppler spectrum fixed (but arbitrary).
Besides being analytically convenient, this approach correctly models different velocities within a given propagation environment.\footnote{The propagation environment determines
the shape of the spectrum while the velocity and the symbol time determine $\fd$.}
We shall henceforth explicitly show the dependence of $\bar{\mathcal{I}}(\cdot)$ and $\bar{\mathcal{I}}^\star(\cdot)$ on $\fd$.
In addition, we recall the notion of spectral shape $\tilde{S}_H(\cdot)$ introduced in (\ref{eze}) and, for the sake of compactness, we introduce the notation
\be
\left[ z \right]_{a}^{b} = \left\{ \begin{array}{ll}
                                         a & z \leq a \\
                                         z & a<z<b \\
                                         b & z \geq b
                                       \end{array}
\right.
\ee

\begin{proposition}
\label{recife}
The optimum pilot overhead for a Rayleigh-faded channel with an arbitrary bandlimited
Doppler spectrum is given by
\bea
\alpha^\star &=&  \left[ \sqrt{(1+\SNR) \, \frac{\dot{C}(\SNR)}{C(\SNR)}\, 2 \fd} \right. \non
&& \left. - \left( (1+\SNR) \, \frac{\ddot{C}(\SNR)}{\dot{C}(\SNR)} + 2 + \frac{1}{2 \, \SNR} \int_{-1}^{+1} \frac{d\xi}{\tilde{S}_H(\xi)} \right) \fd \right]_{2 \fd}^1 \!\!\!\!\!\!\!
+ \mathcal{O}(\fd^{3/2}).
\label{bestalpha}
\eea

{\bf Proof:} See Appendix \ref{barthe}.
\end{proposition}

The expression for $\alpha^\star$ in Proposition \ref{recife} is a simple function involving the perfect-CSI capacity and its derivatives (cf. Section \ref{preliminaries}).
Furthermore,  the leading term in the expansion does not depend on the particular spectral shape.
Only the subsequent term begins to exhibit such dependence, through $\int_{-1}^{+1} d\nu/\tilde{S}_H(\nu) $. For a Clarke-Jakes spectrum, for instance, this integral
equals $\pi^2/2$. For a rectangular spectrum, it equals $4$.

\begin{figure}
  \centering
 \includegraphics[width=0.6\linewidth]{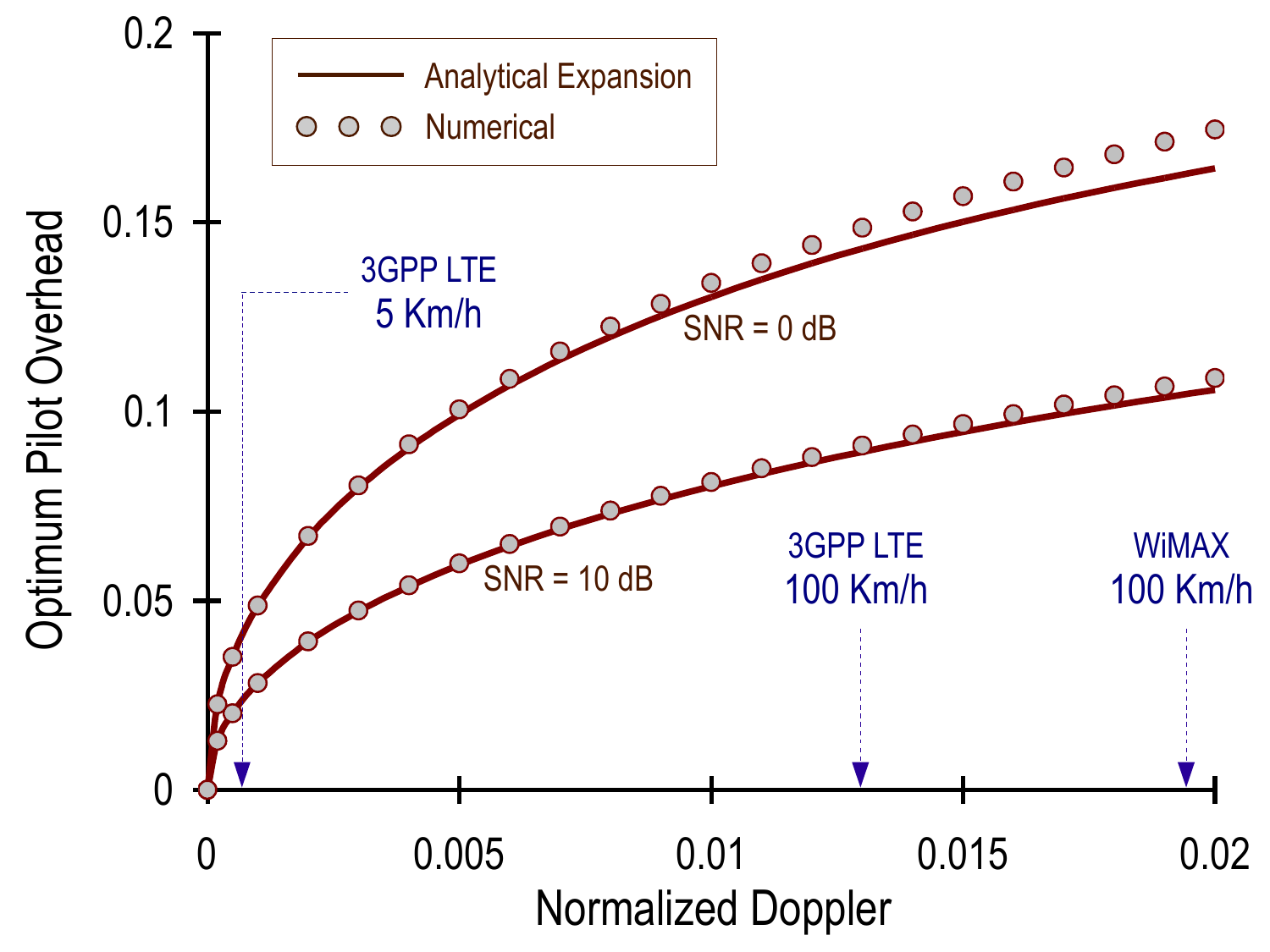}
  \caption{Optimum pilot overhead, $\alpha^\star$, as function of $\fd$ for $\SNR = 0$ dB and $\SNR = 10$ dB with a Clarke-Jakes spectrum.
  Relevant Doppler levels for LTE and WiMAX are highlighted.}
  \label{alpha-doppler}
\end{figure}

\begin{figure}
  \centering
 \includegraphics[width=0.6\linewidth]{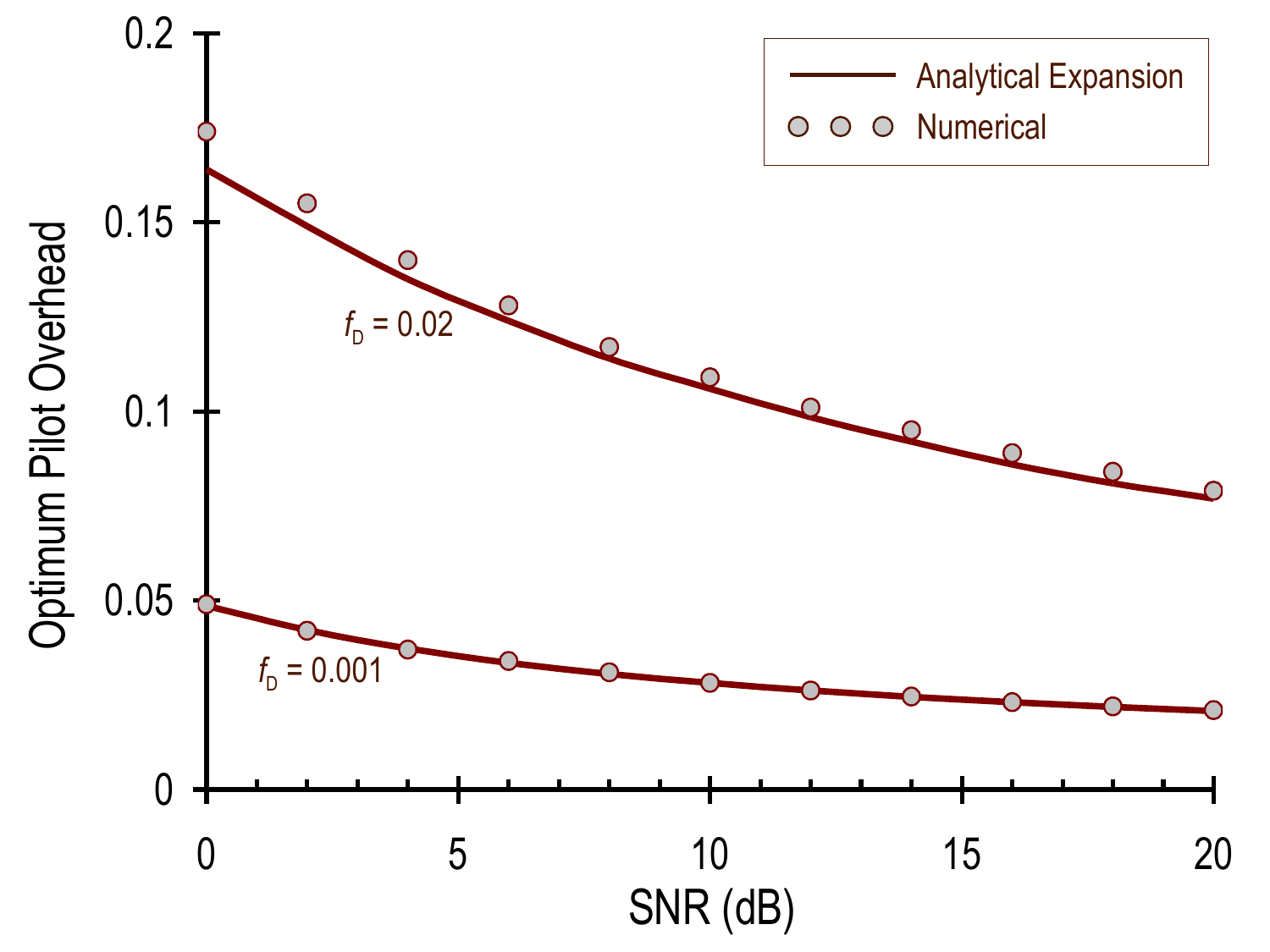}
  \caption{Optimum pilot overhead, $\alpha^\star$, as function of $\SNR$ for $\fd = 0.001$ and $\fd = 0.02$ with a Clarke-Jakes spectrum.}
  \label{alpha-SNR}
\end{figure}

Comparisons between the optimum pilot overhead given by Proposition \ref{recife} and the corresponding
exact value obtained numerically are presented in Figs. \ref{alpha-doppler}--\ref{alpha-SNR}. The agreement is excellent for essentially the entire
range of Doppler and $\SNR$ values of interest in mobile wireless systems.

Once the overhead has been optimized, the corresponding spectral efficiency is given, from (\ref{mallorca}) and Proposition \ref{recife},  by
\be
\label{scuba}
\bar{\mathcal{I}}^\star(\SNR,\fd) = C \left( \SNR \right) - \sqrt{ 8 \fd \left(1 + \SNR \right) C \left( \SNR \right) \dot{C} \left( \SNR \right) } + \mathcal{O}(\fd)
\ee
when $\alpha^\star >  2 \fd$ (up to the order of the expansion). Otherwise,
\be
\bar{\mathcal{I}}^{\star}(\SNR,\fd) = (1 - 2 \fd) \; C \! \left( \frac{\SNR - 1}{2} \right) + \mathcal{O}(\fd^2) .
\ee

\begin{figure}
  \centering
 \includegraphics[width=0.6\linewidth]{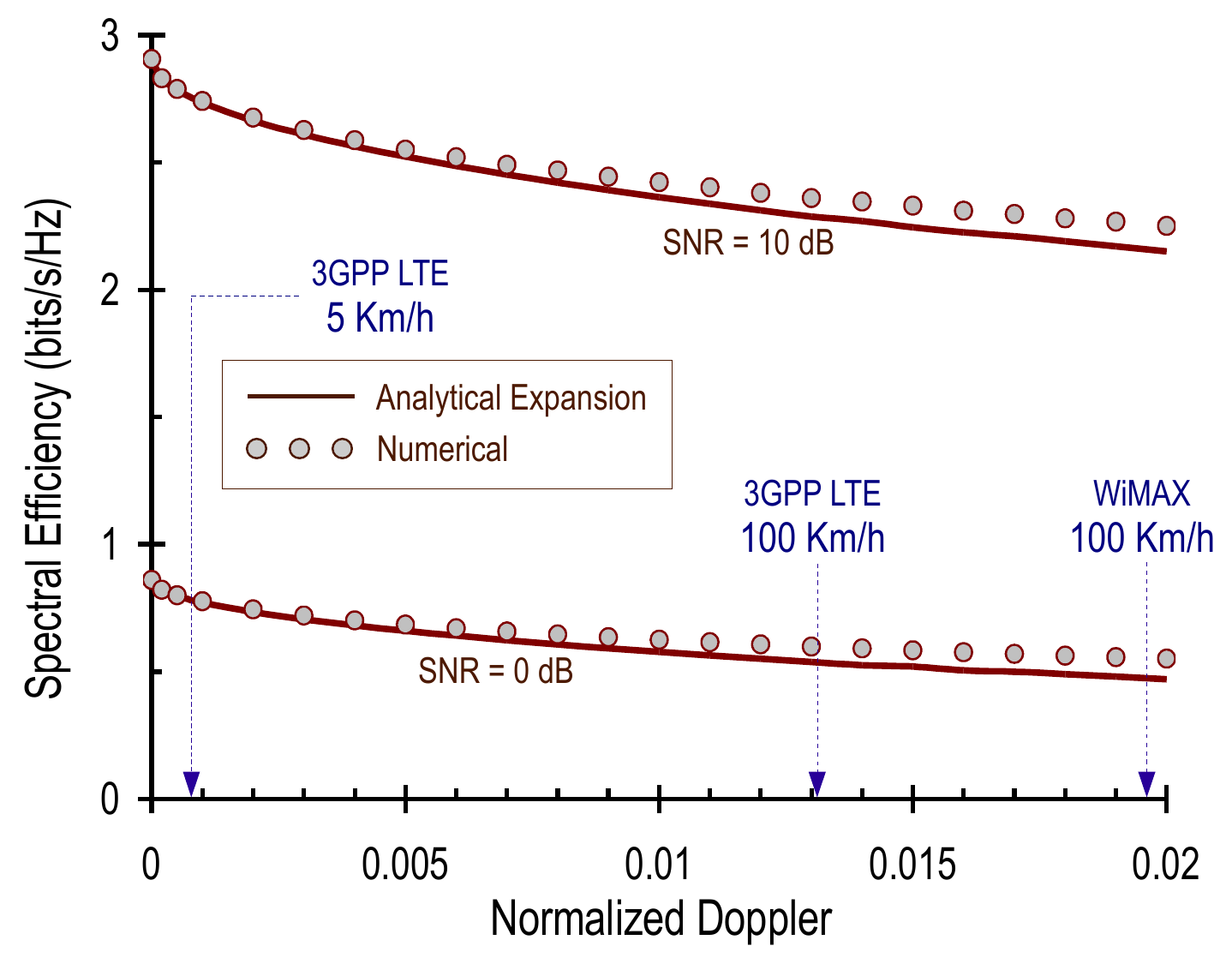}
  \caption{Spectral efficiency with optimum pilot overhead as function of $\fd$  for $\SNR = 10$ dB with a Clarke-Jakes spectrum.
  Relevant normalized Doppler levels for LTE and WiMAX are highlighted.}
  \label{SE-doppler}
\end{figure}

\begin{figure}
  \centering
  \includegraphics[width=0.6\linewidth]{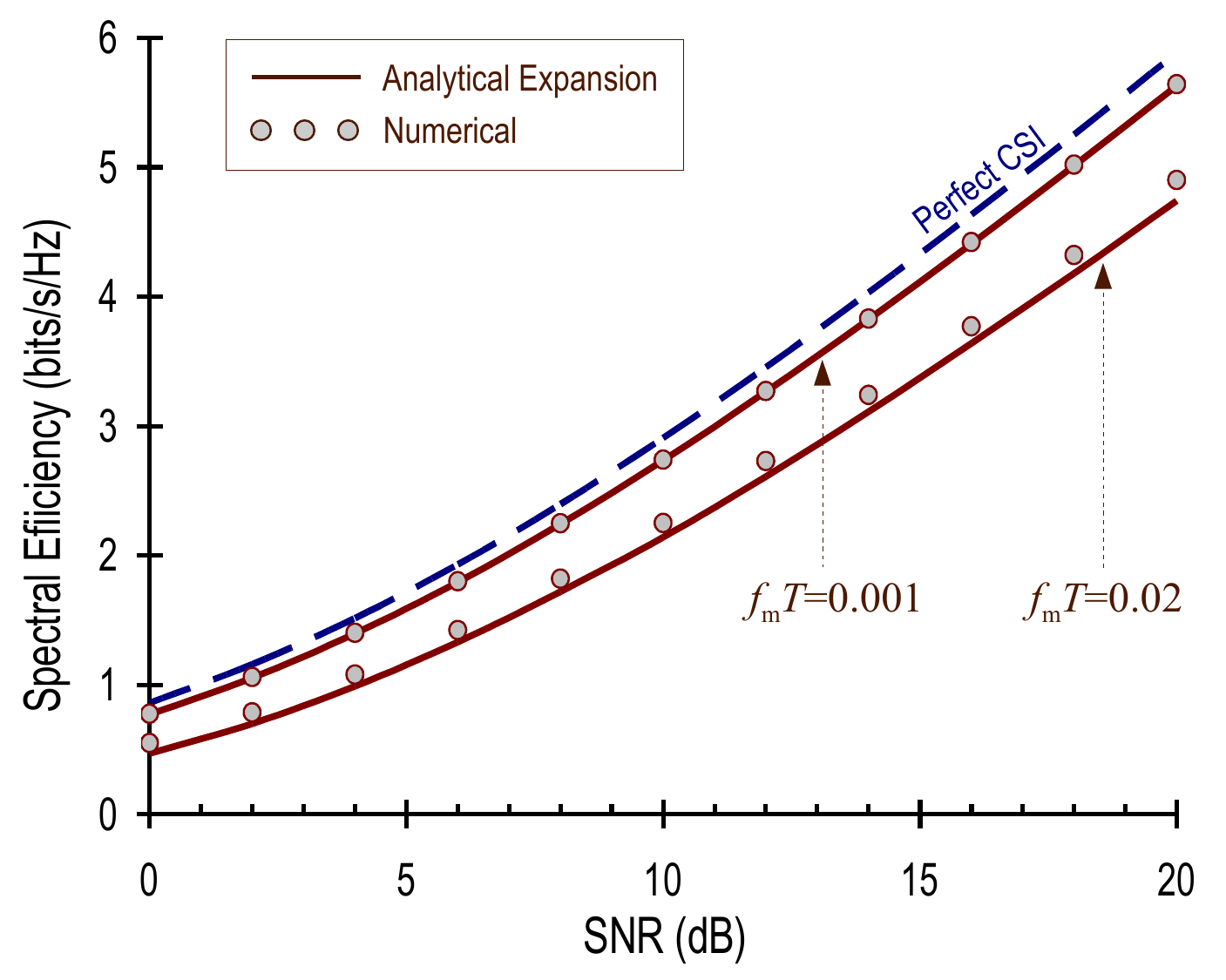}
  \caption{Spectral efficiency with optimum power overhead as function of $\SNR$ for $\fd=0.001$ and $\fd=0.02$ with a Clarke-Jakes spectrum.
  Also shown is the capacity with perfect CSI.}
  \label{SE-SNR}
\end{figure}

As with the optimum overhead, good agreement is shown in Figs. \ref{SE-doppler}--\ref{SE-SNR} between the spectral efficiency in (\ref{scuba})
and its numerical counterpart as rendered by (\ref{mallorca2}).




A direct insight of Proposition \ref{recife} is that the optimum pilot overhead,
$\alpha^*$, and the spectral efficiency penalty w.r.t. the perfect-CSI capacity, $C(\SNR) -
\bar{\mathcal{I}}^\star(\SNR,\fd)$, both depend on the Doppler as $\sqrt{\fd}$.  To gain
an intuitive understanding of this scaling, we can express such penalty for an arbitrary $\alpha$ as (cf. Appendix \ref{barthe}, Eq. \ref{I_taylor2})
\be
\label{I_taylor3}
C(\SNR) - \bar{\mathcal{I}}(\SNR, \alpha, \fd) =  \alpha \, C(\SNR) + \frac{ (1 + \SNR) \dot{C}(\SNR) \, 2 \fd}{\alpha} + \mathcal{O}(\fd).
\ee
The first term in (\ref{I_taylor3}) represents the spectral efficiency loss because only
a fraction $(1 - \alpha)$ of the symbols contain data, while the second term is
the loss on those transmitted data symbols due to the inaccurate CSI.
If $\alpha$ is chosen to be $\mathcal{O}(\fd^s)$ for $0 \leq
s \leq 1$, the first and second terms in (\ref{I_taylor3}) are $\mathcal{O}(\fd^s)$ and
$\mathcal{O}(\fd^{1-s})$, respectively, and thus the overall penalty is
\be
\mathcal{O} \left( \fd^{\min \{ s, 1-s \}} \right) .
\ee
Hence, the spectral efficiency penalty is minimized by
balancing the two terms and selecting $\alpha^\star = \mathcal{O}(\sqrt{\fd})$.

In parsing the dependence of $\alpha^\star$ upon $\SNR$, it is worth noting that
$(1+\SNR) \, \dot{C}(\SNR)/C(\SNR)$ is very well approximated by $1/\log_e (1+ \SNR)$.
Thus, the optimal overhead decreases with $\SNR$ approximately as $1/\sqrt{ \log_e (1 + \SNR) }$.
However, it is important to realize that, although our expansion is remarkably accurate for a wide range of $\SNR$ values,
it becomes less accurate for $\SNR \rightarrow 0$ or $\SNR \rightarrow \infty$.
In fact, in limiting $\SNR$ regimes it is possible to explicitly handle arbitrary Doppler levels \cite{Zheng02,hassibi,DengHaimovich07,lozano08}. Thus, it is precisely for intermediate $\SNR$ values where
the analysis here is both most accurate and most useful, thereby complementing those in the aforegiven references.

\section{Pilot Power Boosting}
\label{milito}

In some systems, it is possible to allocate unequal powers for pilot
and data symbols. Indeed, most emerging wireless systems feature
some degree of pilot power boosting \cite{LTE,wimax}. In our models,
this can be accommodated by defining the signal-to-noise ratios for
pilot and data symbols to be $\rho_{\sf p} \SNR$ and $\rho_{\sf d}
\SNR$, respectively, with \be \label{ran1} \rho_{\sf p} \alpha +
\rho_{\sf d} (1-\alpha) = 1 . \ee Eq. (\ref{mallorca}) continues to
hold, only with \be \label{crowne} \SNR_{\sf eff} = \frac{\SNR \,
(1-\MMSE)}{1/\rho_{\sf d}+ \SNR \cdot \MMSE}.  \ee
The expressions for $\MMSE$ in (\ref{mmse-block}) and (\ref{newmexico}) hold with
$\SNR$ replaced with $\rho_{\sf p} \SNR$.  As a result, with block fading,
\be
\label{jeanbouin}
\MMSE = \frac{1}{1+ \alpha \, n_{\sf b} \, \rho_{\sf p} \SNR}
\ee
while, with continuous fading,
\be
\label{plaza}
\MMSE = 1- \int_{-\fd}^{+\fd} \frac{ \SNR \, S_H^2(\nu)}{1/ (\rho_{\sf p} \alpha ) + \SNR \, S_H(\nu)} \, d\nu.
\ee
It is easily verified, from (\ref{jeanbouin}) and (\ref{plaza}), that the identity between
block fading and continuous fading with a rectangular Doppler spectrum continues to hold under
condition (\ref{veggieworld}).
In turn, for a Clarke-Jakes spectrum, (\ref{plaza}) gives \cite{lozano08}
\be
\MMSE = 1 - \frac{{\rm arctanh} \sqrt{1-\left(\frac{\rho_{\sf p} {\scriptscriptstyle \sf SNR}}{\pi/2} \right)^2}}
{\frac{\pi}{2} \sqrt{\left( \frac{\pi/2}{\rho_{\sf p} {\scriptscriptstyle \sf SNR}} \right)^2-1}} .
\ee

It can be inferred, from (\ref{mallorca}), (\ref{crowne}) and (\ref{plaza}), that it is advantageous to increase $\rho_{\sf p}$ while simultaneously reducing $\alpha$ all the way to $\alpha_{\sf min}$.
Indeed, for the block-fading model, the observation is made in \cite{Zheng02,hassibi,zheng-tse-medard} that, with pilot power boosting, a
single pilot symbol should be inserted on every fading block. With continuous fading, that translates to
\be
\label{eilat}
\alpha =2 \fd
\ee
and the issue is then the optimization of $\rho_{\sf p}$ and $\rho_{\sf d}$.
With $\alpha$ fixed, moreover, the power boosting that maximizes $\bar{\mathcal{I}}(\cdot)$ is directly the one that maximizes $\SNR_{\sf eff}$, i.e.,
\be
\label{molttrist}
\rho_{\sf p}^{\star} = \arg \max_{\rho_{\sf p} \alpha_{\sf min} + \rho_{\sf d} (1-\alpha_{\sf min})=1} \SNR_{\sf
eff} \ee
Although simpler than the optimization in Section \ref{petra}, this nonetheless must
be computed numerically, with the exception of the rectangular spectra/block-fading \cite[Theorem 2]{hassibi}.

As in Section \ref{Kobe}, we circumvent this limitation by expanding the problem in $\fd$.
Again, this yields expressions that are explicit and valid for arbitrary spectral shapes.

\begin{proposition}
\label{spawc}
The optimum power allocation for a Rayleigh-faded channel with an arbitrary bandlimited
Doppler spectrum is given by
\bea
\label{olympics}
\rho_{\sf p}^\star &=& \sqrt{\frac{1 + 1/\SNR}{2 \fd}}  + \mathcal{O}(1) \\
\rho_{\sf d}^\star &=& 1 - \sqrt{\left(1 + \frac{1}{\SNR} \right) 2 \fd } + \mathcal{O}(\fd)
 \label{bestrho} .
\eea

{\bf Proof:} See Appendix \ref{augustrush}.
\end{proposition}
As expected, an order expansion of the
closed-form solution for the rectangular spectra \cite[Theorem 2]{hassibi} matches the above proposition.

As a by-product of Proposition \ref{spawc}, the combination of the expansion of $\SNR_{\sf eff}$ with (\ref{mallorca}) and with (\ref{olympics})--(\ref{bestrho}) leads to
\be
\label{tour}
\bar{\mathcal{I}}^\star(\SNR,\fd) = C \left( \SNR \right) - \sqrt{ 8 \fd \, \SNR \, (1 + \SNR ) } \, \dot{C}( \SNR )  + \mathcal{O}(\fd).
\ee

\begin{figure}
  \centering
 \includegraphics[width=0.6\linewidth]{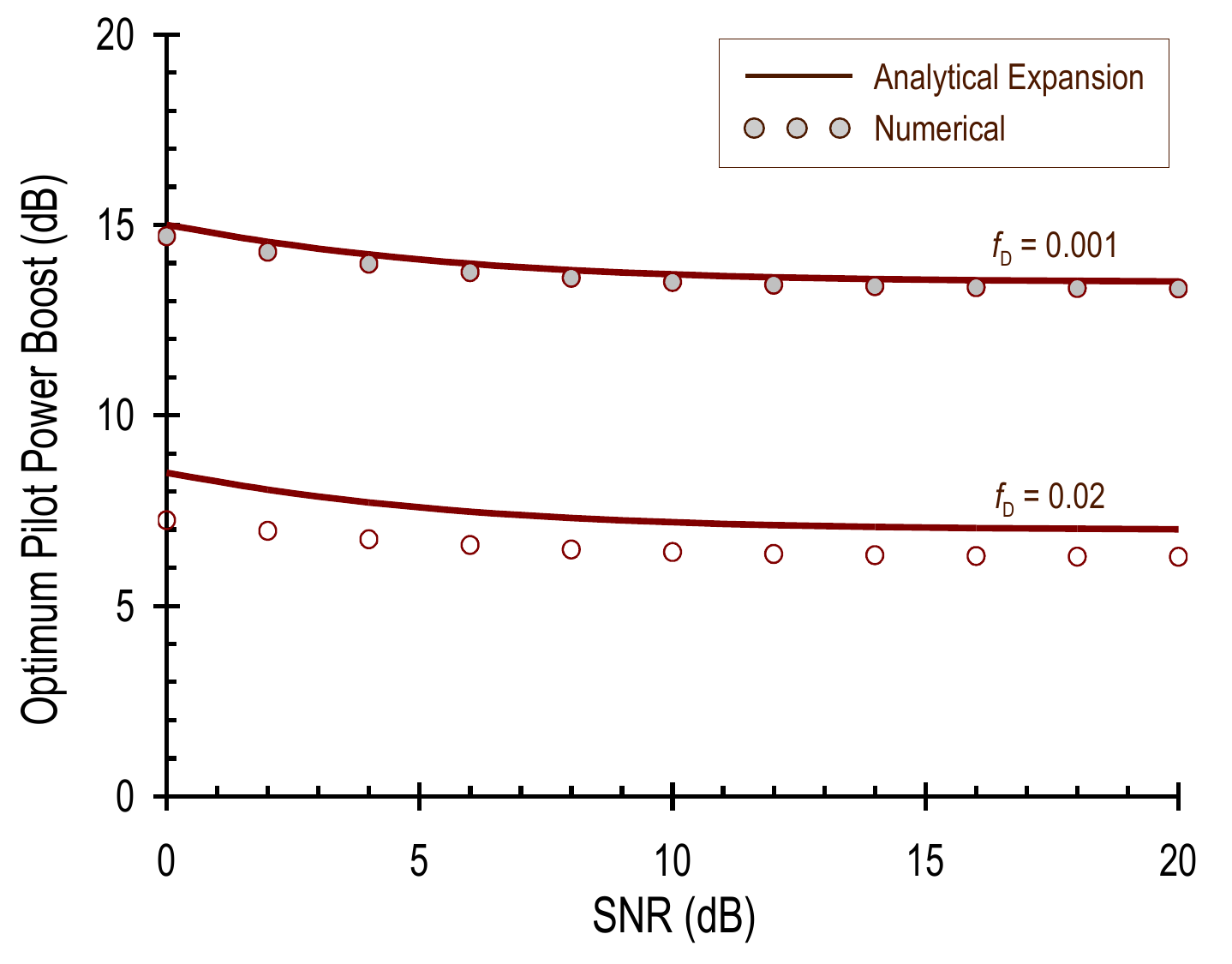}
  \caption{Optimum pilot power boost, $\rho^\star_{\sf p}$, as function of $\SNR$ for $\fd = 0.001$ and $\fd = 0.02$ with a Clarke-Jakes spectrum.}
  \label{rhop-SNR}
\end{figure}

\begin{figure}
  \centering
  \includegraphics[width=0.6\linewidth]{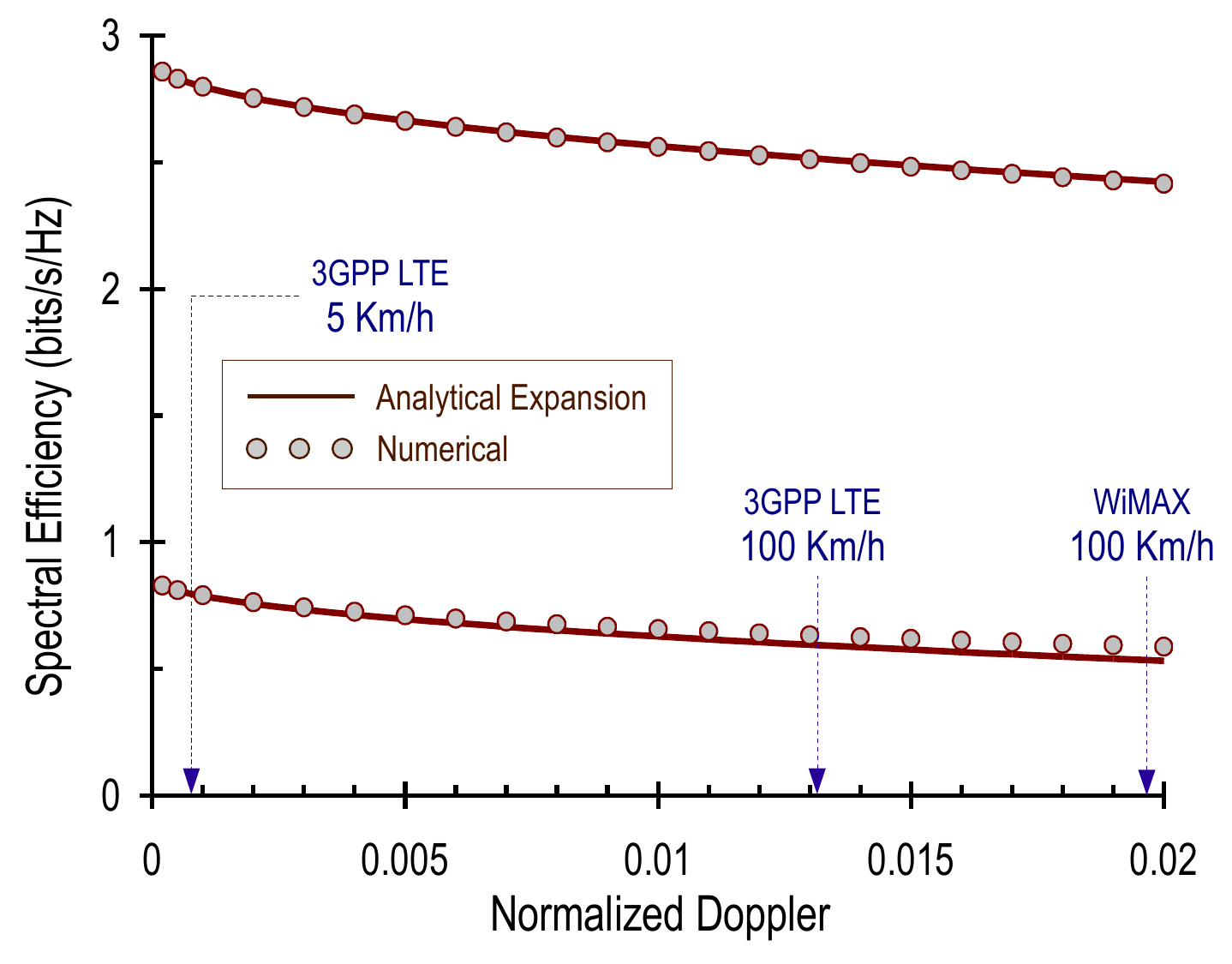}
  \caption{Spectral efficiency with optimum pilot power boost as function of $\fd$  for $\SNR = 10$ dB with a Clarke-Jakes spectrum.
  Relevant normalized Doppler levels for LTE and WiMAX are highlighted.}
  \label{SE-doppler-PowerBoost}
\end{figure}

\begin{figure}
  \centering
  \includegraphics[width=0.6\linewidth]{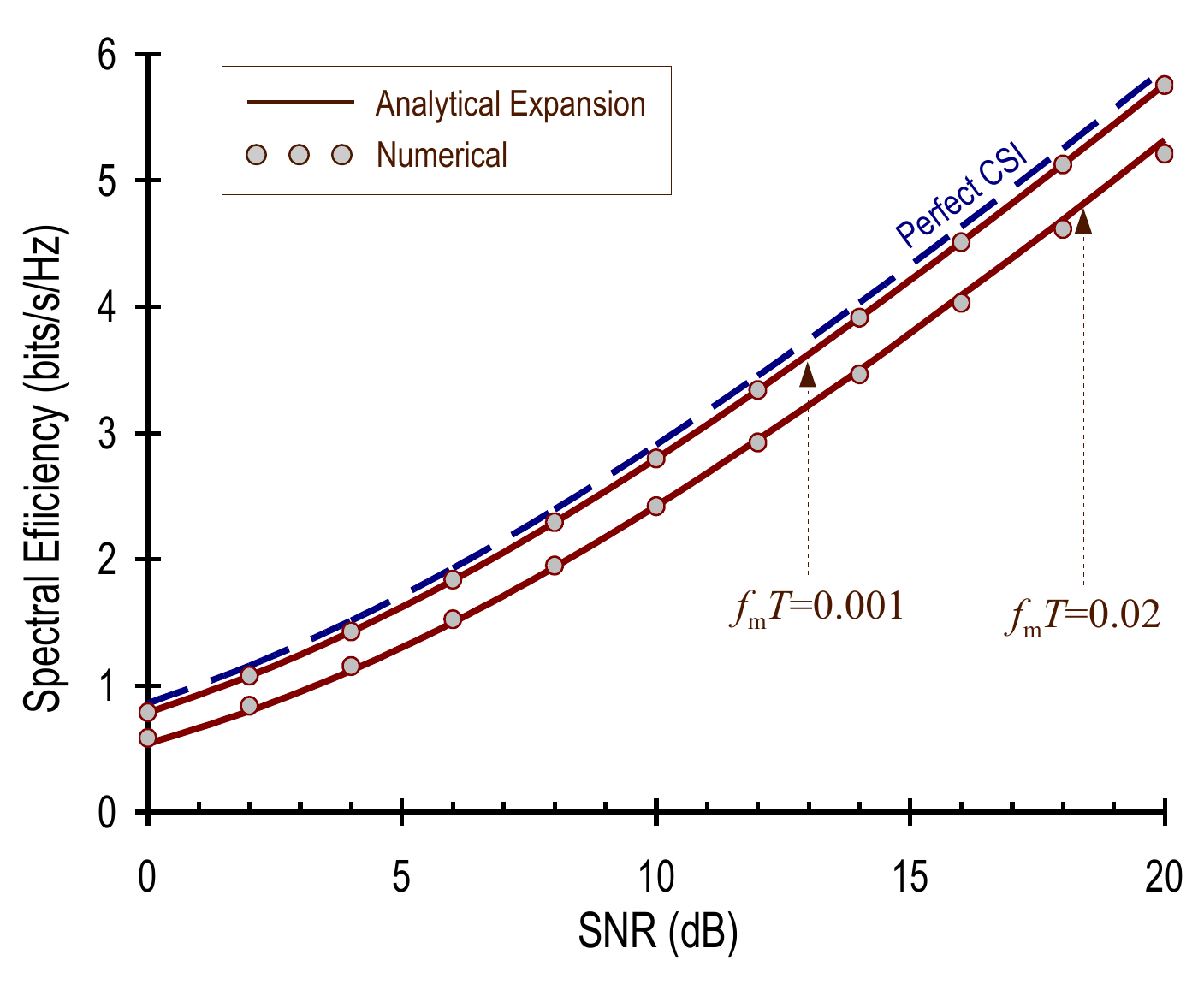}
  \caption{Spectral efficiency with optimum pilot power boost as function of $\SNR$ for $\fd=0.001$ and $\fd=0.02$ with a Clarke-Jakes spectrum.
  Also shown is the capacity with perfect CSI.}
  \label{SE-SNR-PowerBoost}
\end{figure}

A comparison between the optimum pilot power boost given by (\ref{olympics}) and the corresponding value obtained numerically is presented in Fig. \ref{rhop-SNR}. The agreement is excellent.
Good agreement is further shown in Figs. \ref{SE-doppler-PowerBoost}--\ref{SE-SNR-PowerBoost} between the corresponding spectral efficiency in (\ref{tour}) and its exact counterpart, again obtained numerically.


While $\alpha$ is a direct measure of the pilot overhead in terms of bandwidth, the overhead in
terms of power is measured by the product $\rho_{\sf p} \alpha$,
which signifies the fraction of total transmit power devoted to pilot
symbols.
In light of (\ref{eilat}) and Proposition \ref{spawc}, the optimum
pilot power fraction when boosting is allowed equals
\be \label{wadimusa}
\rho_{\sf p}^\star \alpha = \sqrt{\left( 1 +
\frac{1}{\SNR} \right) 2 \fd}  + \mathcal{O}(\fd)  \ee
while without boosting (i.e., with $\rho_{\sf p}=1$) the pilot power fraction
is (from Proposition \ref{recife})
\be
\alpha^\star = \sqrt{(1+\SNR) \, \frac{\dot{C}(\SNR)}{C(\SNR)}\, 2 \fd} + \mathcal{O}(\fd) .
\ee
In both cases the fraction of pilot power fraction is $\mathcal{O} (\sqrt{\fd})$.
Comparing the two, the pilot power fraction with boosting is larger than the fraction
without boosting by a factor
\be
\sqrt{ \frac{C(\SNR)} {\SNR ~ \dot{C}(\SNR)} } .
\ee
This quantity is greater than unity and is increasing in $\SNR$.  Since $\MMSE$ is a decreasing function of $\rho_{\sf p} \alpha$,
this implies that an optimized system with power boosting achieves a smaller $\MMSE$ than one without boosting.

Comparing (\ref{tour}) and (\ref{scuba}), pilot power boosting increases the spectral efficiency by
\be
\sqrt{8 \fd (1+\SNR) \dot{C}(\SNR)} \left( \sqrt{C(\SNR)} - \sqrt{\SNR \, \dot{C}(\SNR)} \right) + \mathcal{O}(\fd)
\ee
which is vanishing for $\SNR \rightarrow 0$ and increases monotonically with $\SNR$.

%
%

\section{Multiantenna Channels}
\label{mimo}

The analysis extends to multiantenna settings in a straightforward manner when there is no antenna correlation.
Letting $\nT$ and $\nR$ denote the number of transmit and receive antennas, respectively,
the channel at time $k$ is now denoted by the $\nR \times \nT$ matrix ${\bf H}(k)$.  Each of the $\nT \nR$ entries of the matrix
varies in an independent manner according to the models described in Section \ref{preliminaries}, for either block or
continuous fading.  The equivalence between block and continuous fading as
per (\ref{veggieworld}) extends to this multiantenna setting, and thus we again restrict our discussion to continuous fading.

We denote the perfect-CSI capacity as
\be
C_{\nT, \nR} (\SNR) = \E \left[ \log_2 \det \left( {\bf I} + \frac{\SNR}{\nT} {\bf H}{\bf H}^{\dagger} \right) \right],
\ee
for which a closed-form expression in terms of the exponential integral can be found in \cite{shinlee}.

The spectral efficiency with pilot-assisted detection now becomes
 \be \label{mallorca-mimo}
 \bar{\mathcal{I}}(\SNR, \alpha) = (1-\alpha) \, C_{\nT, \nR}(\SNR_{\sf eff}) \ee
 with \be
\label{tertulia-mimo} \SNR_{\sf eff} = \frac{\SNR \, (1-\MMSE)}{1+\SNR \cdot \MMSE} \ee where $\MMSE$ is the estimation error for \textit{each} entry
of the channel matrix $\Hm$. This error is minimized by transmitting orthogonal pilot sequences from the various transmit antennas \cite{hassibi}, e.g.,
transmitting a pilot symbol from a single antenna at a time.  A pilot overhead of $\alpha$ thus corresponds to a fraction $\frac{\alpha}{\nT}$ of symbols serving
as pilots for a particular transmit antenna (i.e., for the $\nR$ matrix entries associated with that transmit antenna).
As a result, the per-entry $\MMSE$ is the same as the single-antenna expression in (\ref{northwest})
only with $\alpha$ replaced by $\alpha/\nT$, i.e.,
\be
\MMSE = 1-\int_1^{+1} \frac{\tilde{S}_H^2(\xi)}{\frac{\nT \fd}{\alpha \, \SNR} +\tilde{S}_H(\xi)} \, d\xi.
\label{northwest-mimo}
\ee
This equals the $\MMSE$ for a single-antenna channel with a Doppler frequency of $\nT\fd$. The optimization
w.r.t. $\alpha$ in a multiantenna channel is thus the same as in a single-antenna channel, only with an effective Doppler frequency of $\nT \fd$ and with the function
$C(\SNR)$ replaced by $C_{\nT, \nR} (\SNR)$.  As a result, Proposition \ref{recife} naturally extends into
\bea
\alpha^\star &=&  \left[ \sqrt{(1+\SNR) \,
\frac{\dot{C}_{\nT, \nR}(\SNR)}{C_{\nT, \nR}(\SNR)}\, 2 \nT \fd} \right. \non && \left. - \left( (1+\SNR) \, \frac{\ddot{C}_{\nT,
\nR}(\SNR)}{\dot{C}_{\nT, \nR}(\SNR)} + 2 + \frac{1}{2 \, \SNR} \int_{-1}^{+1} \frac{d\xi}{\tilde{S}_H(\xi)} \right) \nT \fd \right]_{2 \nT \fd}^1
\!\!\!\!\!\!\! + \mathcal{O}(\fd^{3/2}).
\label{bestalpha-mimo}
\eea
Notice here the dependence on  $\sqrt{\nT}$ in the leading term.

When pilot power boosting is allowed, it is again advantageous to reduce $\alpha$ to its minimum value, now given by
$\alpha_{\sf min} = 2 \nT \fd$, and to increase $\rho_{\sf p}$.
In this case the achievable spectral efficiency becomes
\be \label{mallorca-mimo2}
(1-2 \nT \fd) \, C_{\nT, \nR}(\SNR_{\sf eff}) \ee
with $\SNR_{\sf eff}$ as defined in (\ref{tertulia}) and with
\be
\MMSE = 1-\int_1^{+1} \frac{\tilde{S}_H^2(\xi)}{\frac{\nT \fd}{\alpha \, \rho_{\sf p} \, \SNR} +\tilde{S}_H(\xi)} \, d\xi.
\label{northwest-mimo2}
\ee
The optimization of the power boost again corresponds to the maximization of $\SNR_{\sf eff}$  with respect to $\rho_{\sf p}$.
Since $\MMSE$ is the same as for a single-antenna channel with effective Doppler $\nT \fd$, the optimum
pilot power boost for a multiantenna channel with Doppler frequency $\fd$ is exactly the same as the optimum
pilot power boost for a single-antenna channel with the same spectral shape and with Doppler frequency $\nT \fd$.
As a result, the expressions in Section \ref{milito} apply verbatim if $\fd$ is replaced by $\nT \fd$.

Applying (\ref{wadimusa}), the fraction of power devoted to pilots is given by
\be \label{wadimusa-mimo}
\rho_{\sf p}^\star \alpha = \sqrt{\left( 1 +
\frac{1}{\SNR} \right) 2 \nT \fd}  + \mathcal{O}(\fd)  \ee
which increases with $\sqrt{\nT}$.

\begin{figure}
  \centering
  \includegraphics[width=0.6\linewidth]{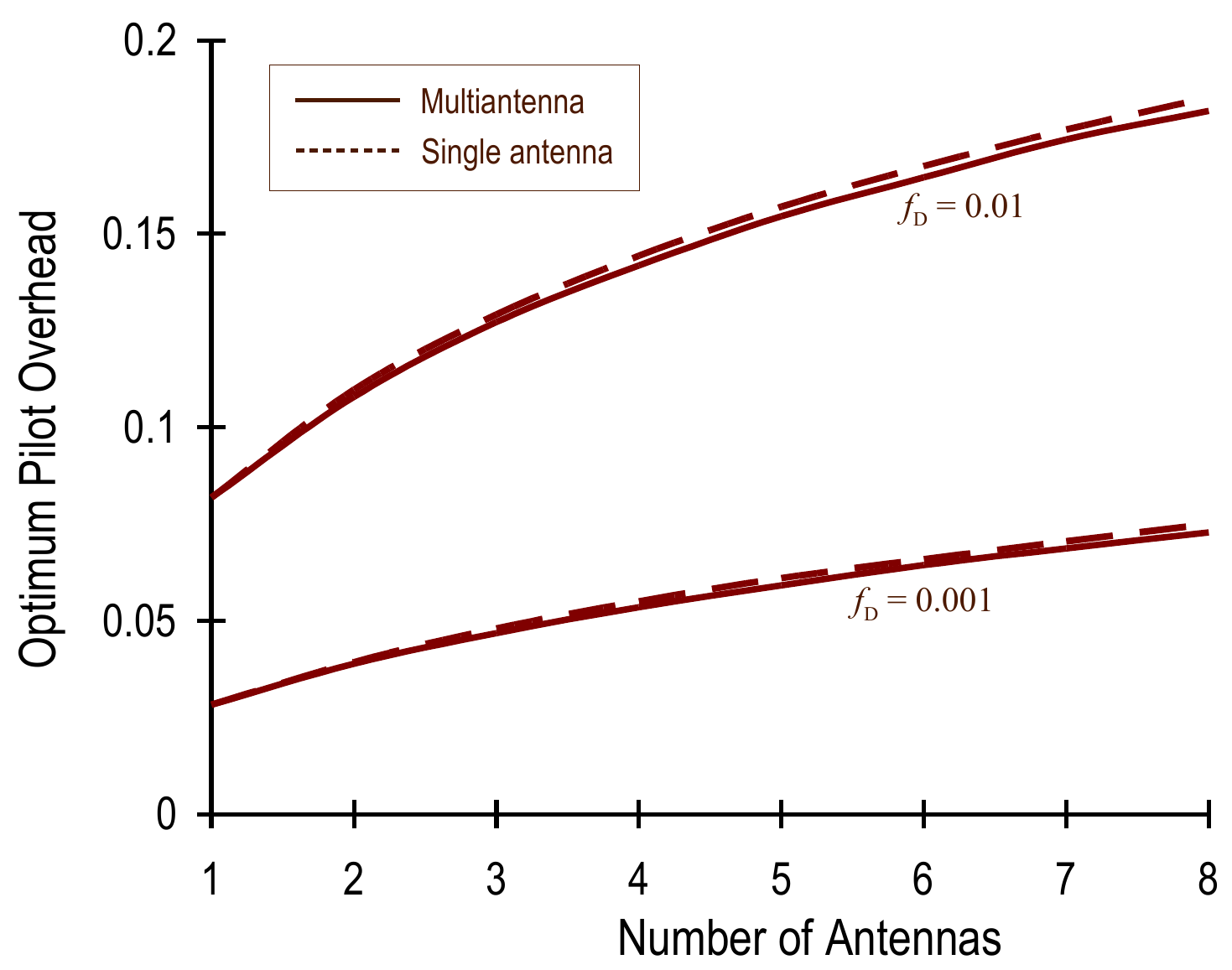}
  \caption{Optimum pilot overhead, $\alpha^\star$, as function of number of antennas ($\nT = \nR$) for $\fd=0.001$ and
  $\fd = 0.01$ for a rectangular spectrum with $\SNR = 10$ dB.  Also shown is the optimal pilot overhead
  for the single-antenna equivalent with a normalized Doppler of $\nT \fd$.}
\label{MIMO}
\end{figure}

Based upon these results, the pilot overhead optimization on a multiantenna channel
with Doppler frequency $\fd$ and a particular spectral shape is effectively equivalent to the optimization on
a single-antenna channel with the same spectral shape and with Doppler frequency $\nT \fd$.  When pilot power boosting is allowed, this
equivalence is in fact exact.  The equivalence is not exact when power boosting is not allowed only because
the perfect-CSI capacity functions $C(\SNR)$ and $C_{\nT, \nR}(\SNR)$ differ.  Roughly speaking, multiple antennas
increase the perfect-CSI capacity by a factor of $\min(\nT,\nR)$ and thus $C_{\nT, \nR}(\SNR) \approx \min(\nT,\nR) C(\SNR)$.
If this approximation were exact,
then the aforementioned equivalence would also be exact.  Although the approximation is not exact,
it is sufficiently valid, particularly for symmetric ($\nT = \nR$) channels, to render the equivalence very
accurate also for the case of non-boosted pilots. To illustrate this accuracy, the optimal pilot overhead
for a symmetric channel at an $\SNR$ of $10$ dB is plotted versus the number of antennas along with the optimal overhead for the single-antenna equivalent
(with Doppler $\nT \fd$) in Fig. \ref{MIMO}. Excellent agreement is seen between the two.

The main implication of the equivalence is that, based upon our
earlier results quantifying the dependence of the pilot overhead on
the Doppler frequency, the optimal overhead (with or without power
boosting) scales with the number of antennas proportional to
$\sqrt{\nT}$.


\section{Summary}

This paper has investigated the problem of pilot overhead optimization in single-user wireless channels.
In the context of earlier work, our primary contributions are two-fold.

First, we were able to unify prior work on continuous- and
block-fading channels and on single- and multiantenna channels: the commonly used block-fading model was shown to
be a special case of the richer set of continuous-fading models in terms of the achievable pilot-based spectral efficiency, and
the pilot overhead optimization for multiantenna chanels is seen to essentially be equivalent to the same optimization for
a single-antenna channel in which the normalized Doppler frequency is multiplied by the number of transmit antennas.

Second, by finding an expansion for the overhead optimization in terms of the fading rate, the square root dependence
of both the overhead and the spectral efficiency penalty was cleanly identified.


\section*{Appendices}
\appendix

\subsection{Proof of Proposition \ref{recife}}
\label{barthe}

We set out to expand $\bar{\mathcal{I}}(\SNR, \alpha)$ w.r.t. $\fd$ about the point $\fd=0$ while holding $\SNR$ and $\alpha$ fixed.
We need
\bea
\frac{\partial \bar{\mathcal{I}}(\SNR,\alpha,\fd)}{\partial \fd} |_{\fd=0} &=&  (1-\alpha) \, \dot{C}(\SNR) \frac{\partial \SNR_{\sf eff}}{\partial \fd} |_{\fd=0} \\
&=& - (1-\alpha) \, \SNR \, (1+\SNR) \, \dot{C}(\SNR) \, \frac{\partial \MMSE}{\partial \fd} |_{\fd=0}
\label{rush}
\eea
and
\bea
\frac{\partial^2 \bar{\mathcal{I}}(\SNR,\alpha)}{\partial \fd^2} |_{\fd=0} &=& (1-\alpha) \left[ \dot{C}(\SNR) \, \frac{\partial^2 \SNR_{\sf eff}}{\partial \fd^2} +
 \ddot{C}(\SNR) \left(\frac{\partial \SNR_{\sf eff}}{\partial \fd}  \right)^2  \right] |_{\fd=0}  \\
&=& - (1-\alpha) \left[ \dot{C}(\SNR) \, \SNR \, (1+\SNR) \left( \frac{\partial^2 \MMSE}{\partial \fd^2}  - 2 \, \SNR \left( \frac{\partial \MMSE}{\partial \fd} \right)^2 \right) \right. \non
&& \left.  +  \ddot{C}(\SNR) \, \SNR^2 (1+\SNR)^2 \left(\frac{\partial \MMSE}{\partial \fd} \right)^2  \right]  |_{\fd=0} .
\label{rush2}
\eea

Based upon (\ref{northwest}), regardless of the shape of the Doppler spectrum we have
\be
\frac{\partial \MMSE}{\partial \fd} |_{\fd=0} = \frac{2}{ \alpha \, \SNR}
\label{turquia}
\ee
where we have used the fact that $\tilde{S}_H(\cdot)$ is bandlimited to $\pm 1$. In turn,
\be
\frac{\partial^2 \MMSE}{\partial \fd^2} |_{\fd=0} = -\frac{2}{(\alpha \, \SNR)^2} \int_{-1}^{+1} \frac{1}{\tilde{S}_H(\nu)} d\nu .
\label{turquia2}
\ee
Combining (\ref{rush}), (\ref{rush2}), (\ref{turquia}) and (\ref{turquia2}),
\bea
\label{I_taylor2}
\bar{\mathcal{I}}(\SNR, \alpha, \fd) & = & (1-\alpha)(1+\SNR) \left[ \frac{C(\SNR)}{1+\SNR} - \dot{C}(\SNR) \frac{2 \fd}{\alpha} \right. \nonumber \\
&& \left. + \left(2 \, (1+\SNR) \, \ddot{C}(\SNR) + \dot{C}(\SNR) \left( \frac{1}{\SNR} \int_{-1}^{+1} \frac{d\xi}{\tilde{S}_H(\xi)} +4 \right) \right) \frac{\fd^2}{\alpha^2}  \right]  + \mathcal{O}(\fd^3)
 \nonumber \\
\eea
which, disregarding the constraints on $\alpha$, is maximized by
\be
\label{euro2008}
\alpha^\star =  \sqrt{2 \fd (1+\SNR) \frac{\dot{C}(\SNR)}{C(\SNR)} }
- \left( (1+\SNR) \frac{\ddot{C}(\SNR)}{\dot{C}(\SNR)} + 2 + \frac{1}{2 \, \SNR} \int_{-1}^{+1} \frac{d\xi}{\tilde{S}_H(\xi)} \right) \fd + \mathcal{O}(\fd^{3/2}).
\ee
To ensure that $\alpha_{\sf min} \leq \alpha^\star \leq 1$ with $\alpha_{\sf min}=2 \fd$, (\ref{euro2008}) must be further constrained as in (\ref{bestalpha}).
Note that the remanent $\mathcal{O}(\fd^{3/2})$, however, is unaffected by the lower constraint (which is $\mathcal{O}(\fd)$). The upper constraint, on the other hand, turns
out to be immaterial.


\subsection{Proof of Proposition \ref{spawc}}
\label{augustrush}

The derivation closely parallels that in Appendix \ref{barthe}. The spectral efficiency equals
\be
\label{lbound}
\bar{\mathcal{I}}(\SNR, \fd) = (1- 2 \fd) \, C \left( \SNR_{\sf eff} \right)
\ee
where the dependence on $\rho_{\sf p}$ and $\rho_{\sf d}$ is concentrated on $\SNR_{\sf eff}$.
To expand $\SNR_{\sf eff}$ w.r.t. $\fd$, we need
\be
\label{danny}
\frac{\partial \SNR_{\sf eff}}{\partial \fd} |_{\fd=0} = - \rho_{\sf d} \SNR \, (1+\rho_{\sf d} \SNR) \, \frac{\partial \MMSE}{\partial \fd} |_{\fd=0} .
\ee
and
\be
\label{danny2}
\frac{\partial^2 \SNR_{\sf eff}}{\partial \fd^2} |_{\fd=0} = - \rho_{\sf d} \SNR \, (1 + \rho_{\sf d} \SNR) \left[  \frac{\partial^2 \MMSE}{\partial \fd^2}
- 2 \rho_{\sf d} \SNR \left( \frac{\partial \MMSE}{\partial \fd}  \right)^2 \right] |_{\fd=0} .
\ee
In order to compute $\partial \MMSE / \partial \fd$ and $\partial^2 \MMSE / \partial \fd^2$, we invoke again the normalized spectral shape in (\ref{eze}) and further use (\ref{ran1}) to rewrite (\ref{plaza}) as
\be
\MMSE = 1-\int_1^{+1} \frac{\tilde{S}_H^2(\xi)}{\frac{\fd}{{\sf \scriptscriptstyle SNR} \, (1-\rho_{\sf d}(1-2 \fd ))} +\tilde{S}_H(\xi)} \, d\xi .
\ee
Then,
\be
\label{layla}
\frac{\partial \MMSE}{\partial \fd} |_{\fd=0} = \frac{2}{\SNR \, (1-\rho_{\sf d})} .
\ee
and
\be
\label{layla2}
\frac{\partial^2 \MMSE}{\partial \fd^2} |_{\fd=0} = - \frac{2}{\SNR (1-\rho_{\sf d})^2} \left( 2 \rho_{\sf d} + \frac{1}{\SNR} \int_{-1}^{+1} \frac{d\nu}{\tilde{S}_H(\nu)} \right) .
\ee
Combining (\ref{danny}), (\ref{danny2}), (\ref{layla}) and (\ref{layla2}), and using the fact that, for $\fd \rightarrow 0$, $\SNR_{\sf eff}$ approaches $\rho_{\sf d} \SNR$, we have
\be
\label{music}
\SNR_{\sf eff} = \rho_{\sf d} \SNR -  \rho_{\sf d} \frac{1+\rho_{\sf d} \SNR}{1 - \rho_{\sf d}} \, 2 \fd
+  \rho_{\sf d} \frac{1+ \rho_{\sf d} \SNR}{(1-\rho_{\sf d})^2} \left( 6 \rho_{\sf d} + \frac{1}{\SNR} \int_{-1}^{+1} \frac{d\nu}{\tilde{S}_H(\nu)} \right) \fd^2
+ \mathcal{O}(\fd^3)
\ee
which, under the constraint that $\rho_{\sf d} \leq 1$, is maximized by
\be
\label{movie}
\rho_{\sf d}^\star = 1- \sqrt{2 \fd \left( 1 + 1/\SNR \right)} + \mathcal{O}(\fd) .
\ee
Analogously, combining (\ref{ran1}) and (\ref{movie}), and with the constraint that $\rho_{\sf d} > 1$,
\be
\label{movie2}
\rho_{\sf p}^\star = \sqrt{\frac{1 + 1/\SNR}{2 \fd}} + \mathcal{O}(1) .
\ee

\clearpage


\end{document}